\documentclass[journal]{IEEEtran}
\IEEEoverridecommandlockouts
\usepackage{amsfonts}
\usepackage{amsmath}
\usepackage{amssymb}
\usepackage{urwchancal}
\usepackage{algorithm}
\usepackage{algorithmicx}
\usepackage{algpseudocode}
\usepackage{dsfont}
\usepackage{authblk}
\usepackage{bbm}
\usepackage{graphicx}
\usepackage{epstopdf}
\usepackage{subfigure}
\usepackage{stfloats}
\usepackage{eqnarray}
\usepackage{makecell}
\usepackage{multirow}
\usepackage{enumerate}
\usepackage{booktabs}
\usepackage{cite}
\usepackage{balance}
\usepackage{color}
\usepackage{bm}
\usepackage{caption}
\usepackage{mathtools}
\DeclarePairedDelimiter\autobracket{(}{)}
\DeclarePairedDelimiter\autosquare{[}{]}
\newcommand{\br}[1]{\autobracket*{#1}}
\newcommand{\sq}[1]{\autosquare*{#1}}
\newcommand{\cb}[1]{\left\{#1\right\}}

\newcommand{\HAV}{\mbox{\tiny HAV}}

\newcommand{\CIR}{\mbox{\tiny CIR}}
\newcommand{\SP}{\mbox{\tiny SP}}

\newcommand{\SC}{\mbox{\tiny SC}}
\newcommand{\SG}{\mbox{\tiny SG}}

\newcommand{\inter}{\mbox{\tiny INTER}}
\newcommand{\intra}{\mbox{\tiny INTRA}}
\newcommand{\act}{\mbox{\tiny ACT}}
\newcommand{\SINR}{\mbox{SINR}}

\renewcommand{\H}{\mbox{\tiny H}}

\newcommand{\T}{{\cal T}}

\newcommand{\C}{\mbox{C}}
\newcommand{\A}{\mbox{\tiny A}}
\newcommand{\F}{\mbox{\tiny F}}
\newcommand{\U}{\mbox{\tiny U}}
\newcommand{\X}{{\cal X}}

\newcommand{\st}{\mbox{s.t.}}
\newcommand{\sumn}{\sum\limits_{n=1}^{N_m}}
\newcommand{\summ}{\sum\limits_{m=1}^M}
\newcommand{\sumk}{\sum\limits_{k=0}^{K-1}}

\newcommand{\sumtk}{\sum\limits_{t_k \in \T_k}}

\newtheorem{remark}{\textbf{Remark}}
\newtheorem{proposition}{\textbf{Proposition}}

\begin{document}
\title{\Large Cross-Layer Scheduling and Beamforming in Smart Grid Powered Small-Cell Networks}


\author[$\ddag$]{\mbox{Yanjie Dong},~\IEEEmembership{Student Member, IEEE}}
\author[$\dag$]{\mbox{Md. Jahangir Hossain},~\IEEEmembership{Senior Member, IEEE}}
\author[$\dag$]{\mbox{Julian Cheng,~\IEEEmembership{Senior Member, IEEE}}}
\author[$\ddag$]{\mbox{Victor C. M. Leung},~\IEEEmembership{Fellow, IEEE}}

\affil[$\ddag$]{Department of Electrical and Computer Engineering\\
The University of British Columbia\\
Vancouver\\
\mbox{BC}\\
Canada}
\affil[$\dag$]{School of Engineering\\
The University of British Columbia\\
Kelowna\\
BC\\
Canada
\authorcr Emails: \{ydong16, vleung\}@ece.ubc.ca\\
\{julian.cheng, jahangir.hossain\}@ubc.ca
\thanks{
This work was supported in part by the National Natural Science Foundation of China under Grant 61671088, in part by a UBC Four-Year Doctoral Fellowship, and in part by the Natural Science and Engineering Research Council of Canada.}
}

\maketitle
\thispagestyle{empty} 
\pagestyle{empty}  
\begin{abstract}
In the small-cell networks (SCNs) with multiple small-cell base stations (ScBSs), the joint design of beamforming vectors, user scheduling and ScBS sleeping is investigated with the constraints on proportional rate.
A long-term grid-energy expenditure minimization problem is formulated for the considered SCNs, which are powered by the smart grid and natural renewable energy.
Since the scheduled user indicators are coupled with the beamforming vectors, the formulated problem is challenging to handle.
In order to decouple the beamforming vectors from the scheduled user indicators, the Lyapunov optimization technique is used.
As a result, a practical two-scale algorithm is proposed to allocate the user scheduling indicators and ScBS sleeping variables at the coarse-grained granularity (frame) as well as obtain the beamforming vectors at the fine-grained granularity (slot).
Numerical results are used to verify the performance of the proposed two-scale algorithm.
\end{abstract}

\section{Introduction}
The wireless data is estimated to exceed $24,220 \times 10^6$ gigabytes per month in 2019, and this amount will be doubled over the next two years \cite{cisco2014}.
The concept of small-cell networks (SCNs) has prevailed as a promising solution in order to cope with the ever-increasing volume of wireless data.
In a typical SCN, the multiple small-cell base stations (ScBSs) connect to the core network via optical fibres or digital subscriber lines, and communicate with the user equipments (UEs) over the radio access links.
Reducing the link distances of transceivers, the ScBSs can decrease the frequency-reuse factor such that the spectrum efficiency of SCN is improved with proper interference mitigation schemes \cite{LiuSept.2018}.
However, the explosion of ScBSs induces the surging energy bills and carbon footprint for the operators of wireless communications \cite{FehskeAug.2011}.
Therefore, the research on reducing energy bills, which is the focus of this work, becomes imperative and attractive for the operators of wireless communications after the ScBSs are deployed.

A quantitative study estimated that the radio access links will consume around 29\% energy of SCNs \cite{FehskeAug.2011, WuAug.2017}.
Hence, the first research direction to reduce the energy bills focuses on reducing the energy consumption \cite{DahroujMay2010, LakshminarayanaOct.2015} or increasing the energy efficiency \cite{HeJune2014} of the ScBSs.
For example, the energy consumption minimization problems with short-term and long-term communication quality of service (QoS) were respectively studied in \cite{DahroujMay2010} and \cite{LakshminarayanaOct.2015} via downlink beamforming of multiple ScBSs in the SCNs.
The second research direction leverages the paradigm shift from the traditional power grid to smart grid \cite{BuAug.2012, FarooqFeb.2017, DongDec.2017}.
With the two-way energy trading between the smart grid and SCNs, incorporating the natural renewable energy (NRE) into the SCNs becomes an ecologically- and economically-friendly solution to cut down the energy bills.
Due to the volatility of NRE arrival, several research efforts have been made in the design of smart grid powered cellular communication frameworks \cite{BuAug.2012, FarooqFeb.2017} and algorithms \cite{DongDec.2017, WangMay2016, HuangJune2017, GuoJuly2018}.
For example, the authors in \cite{DongDec.2017} investigated the impact of volatility of NRE arrival on the packet rates.
By formulating a long-term grid-energy (LTGE) expenditure minimization problem, the authors in \cite{DongDec.2017} revealed that the LTGE expenditure can be reduced by sacrificing the system packet rate.
The authors in \cite{WangMay2016} studied the long-term data-rate maximization problem with a constraint on the LTGE expenditure in the smart grid powered communications.
Using the dirty paper coding at the multiple-input-multiple-output base station, the authors in \cite{WangMay2016} proposed an online beamforming algorithm and proved the asymptotical optimality.
While the proposed algorithms in \cite{DongDec.2017, WangMay2016} are only applicable to the single-cell scenario, the authors in \cite{HuangJune2017} considered the NRE coordination and base station load control problem in a multicell scenario.
The authors in \cite{GuoJuly2018} investigated the joint content delivery and UE association in a scenario with multiple ScBSs.
However, the current works are based on a common assumption that all the resource allocation actions are performed with a single granularity.
In other words, all the resources are allocated over slots.
Due to the volatility of NRE arrival, several resources (e.g., ScBS sleeping variables and scheduled UE indicators) need to be reallocated at a coarse-grained granularity in practical systems since frequent opening/closing the ScBSs or UE induces issue of reliability.
Limited literature has investigated the two-scale resource allocation schemes.
The authors in \cite{WangMay2018} proposed the dynamic beamforming and grid-energy merchandizing algorithm to minimize the LTGE expenditure in a single cell scenario.
The authors in \cite{YuApr.2016} investigated the joint network selection, subchannel and power allocation in the integrated cellular and Wi-Fi networks.
However, the aforementioned literatures \cite{DongDec.2017, WangMay2016, HuangJune2017, GuoJuly2018, YuApr.2016, WangMay2018} do not consider issues on fairness and scheduling among of the UEs.

Different from \cite{WangMay2018, YuApr.2016}, we minimize the LTGE expenditure via the joint design of beamforming vectors, scheduled UE indicators and ScBS sleeping variables among multiple ScBSs in a two-scale framework.
Moreover, we guarantee fairness among UEs according to the backlog of traffic queues of UEs.
The contributions of this work are summarized as follows.
First, we investigate the LTGE expenditure minimization problem in the SCN via the joint design of beamforming vectors, scheduled UE indicators and ScBS sleeping variables in the SCN with multiple ScBSs.
The design of beamforming vectors belong to physical layer, and the design of scheduled UE indicators and ScBS sleeping variables belong to the upper layers.
Hence, the investigated LTGE expenditure minimization is a cross-layer problem.
Second, we theoretically prove that the decisions on scheduled UE indicators and ScBS sleeping variables depend on the accumulated effect of channel quality.
Moreover, we also reveal that the proposed algorithm can approach the optimal LTGE expenditure via tuning a control parameter. Numerical results are used to verify the performance of our proposed algorithm.

%

\vspace{-0.35 cm}
\section{System Model and Problem Formulation}
\vspace{-0.2 cm}
We consider an SCN with $M$ ScBSs.
The $m$-th ScBS is associated with $N_m$ UEs.
Each ScBS is equipped with $N_T$ transmit antennas, and each UE is equipped with single antenna.
Moreover, each ScBS connects to the core network (CN) and UEs via the optical-fibre link and wireless links, respectively.
Each ScBS is powered by NRE and smart grid.
Since the NRE arrival rates and channel coefficient vectors vary at different time scales in practice \cite{LiApr.2015}, a two-scale framework is considered.
Let each frame consist of $T$ slots.
The average NRE arrival rates vary over frames, and the channel coefficient vectors vary over slots.
We respectively denote the indices for frame and slot as the $k$-th frame and the $t_k$-th slot with $k = 0, 1, \ldots, \infty$ and $t_k \in \T_k \triangleq \cb{t_k \left| kT\le t_k \le \br{k+1}T - 1\right.}$.
Moreover, each slot has unit duration; therefore, we use the terms ``energy'' and ``power'' interchangeably at the scale of slot.

\vspace{-0.3 cm}
\subsection{Traffic Model}
%
%
%

\subsubsection{Access Queue}
We consider that the $m$-th ScBS maintains $N_m$ access queues for the associated UEs, and dynamic equation for the $n$-th access queue of the $m$-th ScBS (or the $\br{m,n}$-th access queue) is given as
\begin{equation}\label{eqa:02}
q_{m,n}^{\A}\br{t_k+1} = q_{m,n}^{\A}\br{t_k} - r_{m, n}\br{t_k} + \nu_{m, n}\br{t_k}
\end{equation}
where $q_{m,n}^{\A}\br{t_k+1}$ and $q_{m,n}^{\A}\br{t_k}$ are the backlogs of the $\br{m,n}$-th access queue at the beginning of the $\br{t_k +1}$-th and the \mbox{$t_k$-th} slot, respectively;
$\nu_{m,n}\br{t_k}$ and $r_{m,n}\br{t_k}$ are, respectively, the traffic arrival rate and service rate of the $\br{m,n}$-th access queue at the $t_k$-th slot.
Here, the value of $\nu_{m,n}\br{t_k}$ is set as
\begin{equation}\label{eqa:03}
\nu_{m,n}\br{t_k} = \left\{ \begin{array}{l}
\lambda_{m,n}, t_k = kT \\
0, \mbox{otherwise.}
\end{array} \right.
\end{equation}

\subsubsection{Processing Queue}
We consider that the $n$-th UE of the $m$-th ScBS (or the $\br{m,n}$-th UE) maintains a processing queue (or the $\br{m,n}$-th processing queue) for the upper layer processing.
The dynamic equation for the $\br{m,n}$-th processing queue is given as
\begin{equation}\label{eqa:04}
q_{m,n}^{\U}\br{t_k + 1} = q_{m,n}^{\U}\br{t_k} - s_{m,n}\br{t_k} + r_{m,n}\br{t_k}
\end{equation}
where $q_{m,n}^{\U}\br{t_k + 1}$ and $q_{m,n}^{\U}\br{t_k}$ are the backlogs at the beginning of the $\br{t_k+1}$-th and the $t_k$-th slot.
We consider the constant service rate of the $\br{m,n}$-th processing queue.
Therefore, the service rate $s_{m,n}\br{t_k} \triangleq \min\br{\bar s_{m,n}, q_{m,n}^{\U}\br{t_k}}$ where $\bar s_{m,n}$ denotes the average service rate of the $\br{m,n}$-th processing queue.

\vspace{-0.3 cm}
\subsection{Signal Model}
Let $\bm h_{m,n}\br{t_k}$ be the channel coefficient vector of the link between the $n$-th UE and the $m$-th ScBS (or $\br{m,n}$-th access link) at the $t_k$-th slot.
Here, $\bm h_{m,n}\br{t_k} \sim {\cal CN}\br{\bm 0, d_{m,n}^{-\chi}\bm I_{N_T}}$ follows circularly symmetric complex Gaussian distribution (CSCG) where $d_{m,n}$ and $\chi$ are, respectively,  the distance of the $\br{m,n}$-th access link and pathloss exponent \cite{DongMay2017}.

Let $a_{m,n}\sq{k}$ be the scheduled UE indicator with $a_{m,n}\sq{k} = 1$ denoting that the $\br{m,n}$-th UE is scheduled at the \mbox{$k$-th} frame; otherwise, $a_{m,n}\sq{k} = 0$.
Therefore, the received signal and signal-to-interference-plus-noise ratio (SINR) of the $\br{m,n}$-th UE at the $t_k$-th slot are, respectively, denoted as
\begin{multline}\label{eqa:05}
y_{m, n}\br{t_k} =  \sqrt{a_{m,n}\sq{k}} \bm h_{m,n}^{\H}\br{t_k}\bm w_{m,n}\br{t_k}\\
+\sum\limits_{i\neq n} \sqrt{a_{m,i}\sq{k}} \bm h_{m,n}^{\H}\br{t_k}\bm w_{m,i}\br{t_k} \\
+ \sum\limits_{j \neq m}\sum\limits_{i = 1}^{N_j} \sqrt{ a_{j,n}\sq{k} } \bm h_{j,n}^{\H}\br{t_k}\bm w_{j,i}\br{t_k}  + z_{m,n}\br{t_k}
\end{multline}
and
\begin{equation}\label{eqa:06}
\SINR_{m,n}\br{t_k} = \frac{ a_{m,n}\sq{k} \left|\bm h_{m,n}^{\H}\br{t_k}\bm w_{m,n}\br{t_k}\right|^2}{I_{m, n}^{\intra}\br{t_k} + I_{m, n}^{\inter}\br{t_k} + \sigma_{m,n}^2}
\end{equation}
where the term $z_{m,n}\br{t_k} \sim {\cal CN}\br{0, \sigma_{m,n}^2}$ is the additive white Gaussian noise (AWGN) of the $\br{m, n}$-th UE at the $t_k$-th slot;
$\bm w_{m,n}\br{t_k}$ denotes the single-stream beamforming vector for the $\br{m,n}$-th UE at the $t_k$-th slot;
and the intra-cell interference and inter-cell interference terms are, respectively, given as
\begin{equation}\label{eqa:07}
I_{m,n}^{\intra}\br{t_k} = \sum\limits_{i\neq n}  {a_{m,i}\sq{k}} \left|\bm h_{m,n}^{\H}\br{t_k}\bm w_{m,i}\br{t_k}\right|^2
\end{equation}
and
\begin{equation}\label{eqa:08}
I_{m,n}^{\inter}\br{t_k} = \sum\limits_{j \neq m}\sum\limits_{i = 1}^{N_j} {a_{j,i}\sq{k}} \left|\bm h_{j,n}^{\H}\br{t_k}\bm w_{j,i}\br{t_k}\right|^2.
\end{equation}
Hence, the data rate of the $\br{m,n}$-th access queue at the $t_k$-th slot is given as
$r_{m,n}\br{t_k} = \log\br{1 + \SINR_{m,n}\br{t_k}}$.

Based on \eqref{eqa:05}, the consumed power of the $m$-th ScBS is denoted as
\begin{equation}\label{eqa:09}
\begin{split}
&P^{\SC}_m\br{t_k} \\
=& \left\{ \begin{array}{l}
\frac{1}{\eta} \sum\limits_{n \in {\cal N}_m^{\act}\sq{k}} \left\|\bm w_{m,n}\br{t_k}\right\|_{\F}^2 +  P_m^{\CIR}, \sumn a_{m,n}\sq{k} > 0 \\
0, \sumn a_{m,n}\sq{k} = 0
\end{array} \right.
\end{split}
\end{equation}
where $P_m^{\CIR} \triangleq P_m^{\SP}\br{0.87 + 0.1N_T + 0.03 N_T^2}$ \cite{DongDec.2017} with $P_m^{\SP}$ as the consumed power on baseband processing of the \mbox{$m$-th} ScBS; and $\eta$ is the power amplifier efficiency of the \mbox{$m$-th} ScBS.
Here, ${\cal N}_m^{\act}\sq{k}$ denotes the set of scheduled UEs of the $m$-th ScBS at the $k$-th frame.


\subsection{Model of Grid-Energy Expenditure}
The grid-energy expenditure of SCN at the $k$-th frame is obtained as
\begin{equation}\label{eqa:11}
\begin{split}
G\sq{k} &= \sumtk \br{ \alpha_b\br{  P^{\SG}\br{t_k} }^+ - \alpha_s\br{ - P^{\SG}\br{t_k} }^+ } \\
&= \sumtk \br{ \br{\alpha_b - \alpha_s}\br{  P^{\SG}\br{t_k} }^+ + \alpha_s P^{\SG}\br{t_k} }
\end{split}
\end{equation}
where $\alpha_b$ and $\alpha_s$ are the electricity prices for purchasing from and selling to the smart grid with $\alpha_b > \alpha_s$ to avoid the redundant energy trading; $P^{\SG}\br{t_k}$ denotes the amount of two-way energy trading between the smart grid and ScBSs at the $t_k$-th slot.
Here, $P^{\SG}\br{t_k}$ takes positive (or negative) value when purchasing from (or selling to) the smart grid.

\begin{remark}
By formulating the grid-energy expenditure of SCN as in \eqref{eqa:11}, we consider an ideal case where one ScBS can trade NRE with other ScBSs free of charge.
The non-ideal case will be considered in the extended version of the conference paper.
\end{remark}

\subsection{Problem Formulation}
Our objective is to minimize the grid-energy expenditure via jointly designing the scheduled UE indicators $\{a_{m,n}\sq{k}\}_{\forall m, n, k}$ in each frame as well as the beamforming vectors and amount of trading energy $\{\bm w_{m,n}\br{t_k}, P^{\SG}\br{t_k}, \}_{\forall m,n,k,t_k}$ in each slot.
Due to the lack of knowledge on stochastic arrival of NRE and variations of CSI, we formulate the LTGE expenditure minimization problem with the following constraints:
\begin{itemize}
  \item Rate-limit constraints:
  \begin{equation}\label{eqa:12}
  r_{m, n}\br{t_k} \le q_{m,n}^{\A}\br{t_k}, \forall m,n
  \end{equation}
  which guarantee that each ScBS does not transmit blank information.
  \item Proportional-rate constraints:
  \begin{equation}\label{eqa:13}
  \frac{r_{m, n}\br{t_k}}{ r_{j, i}\br{t_k} } = \frac{\psi_{m,n}}{\psi_{j,i}}, n \in {\cal N}_m^{\act}\sq{k}, i \in {\cal N}_j^{\act}\sq{k}, \forall m, j.
  \end{equation}
  For example, $\frac{\psi_{m,n}}{\psi_{j,i}} = \frac{q^{\A}_{m,n}\br{t_k}}{q^{\A}_{j,i}\br{t_k}}$ guarantees that the UE with larger backlog obtains better service rate at each slot.
  \item Slot-level power constraints:
  \begin{equation}\label{eqa:14}
  \sum\limits_{n \in {\cal N}_{m}^{\act}\sq{k}}  \left\|\bm w_{m,n}\br{t_k}\right\|_{\F}^2 \le P_m^{\max}, \forall m
  \end{equation}
  where $P^{\max}_m$ is the maximum transmit power of the $m$-th ScBS.
  \item Power balance constraint:
  \begin{equation}\label{eqa:15}
  P^{\SG}\br{t_k} + \summ \frac{{E}_m^{\HAV}\sq{k}}{T} = \summ P_m^{\SC}\br{t_k}
  \end{equation}
  which guarantees that the consumed grid energy is balanced with the harvested NRE and merchandized energy with smart grid at the $t_k$-th slot.
  \item Queue-stable constraints:
  \begin{align}
  &\limsup\limits_{K\rightarrow \infty}\frac{1}{K} \label{eqa:16}\\
  & \times\sumk \mathds{E}_{\cal X}\cb{ q_{m,n}^{\A}\sq{k} +  q_{m,n}^{\U}\sq{k}} < \infty, \forall m, n \nonumber
  \end{align}
  where $q_{m,n}^{\A}\sq{k+1}\triangleq q_{m,n}^{\A}\br{t_k}\left|_{t_k = \br{k+1}T}\right. $ and $q_{m,n}^{\A}\sq{k} \triangleq q_{m,n}^{\A}\br{t_k}\left|_{t_k = kT}\right.$.
  The queue-stable constraints indicate that the data of UEs will be served in finite time.
\end{itemize}
Here, the operator $\mathds{E}_{\cal X}\cb{\cdot}$ denotes the expectation over the random sources ${\cal X} \triangleq \cb{\bm h_{m,n}\br{t_k}, E_m^{\HAV}\sq{k}}_{\forall m, n, k, t_k}$.

As a result, the LTGE expenditure minimization problem is formulated as
\begin{subequations}\label{eqa:17}
\begin{align}
\min\limits_{\cal Y} & \lim\limits_{K \rightarrow \infty} \frac{1}{K}\sumk \mathds{E}_{\cal X}\cb{ G\sq{k} } \label{eqa:17a}\\
\st\; &  \eqref{eqa:12}-\eqref{eqa:16}
\end{align}
\end{subequations}
where resource-allocation-variable set is defined as ${\cal Y} \triangleq \{\bm w_{m,n}\br{t_k}, P^{\SG}\br{t_k},  a_{m,n}\sq{k}\}_{\forall m, n, k, t_k}$.

Note that the LTGE expenditure minimization problem \eqref{eqa:17} is challenging to handle via classical convex optimization methods.
Since the scheduled UE indicators are coupled with the beamforming vectors, we are motivated to use the Lyapunov optimization method to obtain a feasible solution to the LTGE expenditure minimization problem \eqref{eqa:17}.
Moreover, we also demonstrate that the LTGE expenditure is asymptotically minimized by sacrificing the end-to-end delay of UEs when scheduling UEs and switching on/off ScBSs are considered.

\section{Joint Beamforming, UE Scheduling and ScBS Sleeping}
We define the Lyapunov function of LTGE expenditure minimization problem as
\begin{equation}\label{eqa:18}
  L\sq{k}
= \frac{1}{2} \summ\sumn \br{ \br{ q_{m,n}^{\A}\sq{k} }^2 + \br{q_{m,n}^{\U}\sq{k}}^2 }
\end{equation}
where $q_{m,n}^{\A}\sq{k} \triangleq q_{m, n}^{\A}\br{kT}$ and $q_{m,n}^{\U}\sq{k} \triangleq q_{m, n}^{\U}\br{kT}$.
Then, we introduce the one-frame drift function as \cite{Neelybook}
\begin{equation}\label{eqa:19}
\Delta_{\X} \triangleq \mathds{E}_{\X} \cb{ L\sq{k + 1} -  L\sq{k}}.
\end{equation}
Thus, we obtain the one-frame Lyapunov \mbox{drift-plus-penalty} function as \cite{Neelybook}
\begin{equation}\label{eqa:20}
\Delta_{\X} + V\mathds{E}_{\cal X}\cb{ G\sq{k} }.
\end{equation}

\begin{proposition}\label{pr:01}
The one-frame Lyapunov drift-plus-penalty function in \eqref{eqa:20} is upper-bounded as
\begin{align}
    & \Delta_{\X} + V \mathds{E}_{\cal X}\cb{ G\sq{k} }  \label{eqa:21}\\
\le & \Psi + V \mathds{E}_{\cal X}\cb{ G\sq{k} } \nonumber \\
& + \summ\sumn q_{m,n}^{\A}\sq{k}f_{m,n}^{\A}\sq{k} + \summ\sumn q_{m,n}^{\U}\sq{k}f_{m,n}^{\U}\sq{k}  \nonumber
\end{align}
where the functions $f_{m,n}^{\A}\sq{k}$ and $f_{m,n}^{\U}\sq{k}$ are, respectively, defined as
\vspace{-0.3 cm}
\begin{equation}\label{eqa:23}
f_{m,n}^{\A}\sq{k} \triangleq \mathds{E}_{\cal X}\cb{ \lambda_{m,n} - \sumtk r_{m,n}\br{t_k} }
\end{equation}
\vspace{-0.3 cm}
and
\vspace{-0.1 cm}
\begin{equation}\label{eqa:24}
f_{m,n}^{\U}\sq{k} \triangleq \mathds{E}_{\cal X} \cb{ \sumtk\br{r_{m,n}\br{t_k} - s_{m,n}\br{t_k}} }.
\end{equation}
\vspace{-0.3 cm}
\end{proposition}
\begin{IEEEproof}
See Appendix.
\end{IEEEproof}

Minimizing the right-hand side (RHS) of \eqref{eqa:21} subject to constraints in \eqref{eqa:12}--\eqref{eqa:16} results in a feasible solution to the LTGE expenditure minimization problem \eqref{eqa:17}.

\vspace{-0.3 cm}
\subsection{UE Scheduling Analysis}
After some algebraic manipulations on RHS of \eqref{eqa:21}, we obtain the term related to $\cb{ r_{m,n}\br{t_k}}_{\forall m,n,k,t_k}$ as
\begin{equation}\label{eqa:26}
\summ\sumn \br{ q_{m,n}^{\U}\sq{k} - q_{m,n}^{\A}\sq{k} } \mathds{E}_{\cal X}\cb{\sumtk  r_{m,n}\br{t_k} }.
\end{equation}

Here, the data rate of the $\br{m,n}$-th UE is coupled with the scheduled UE indicator.
Since our objective is to minimize the term in \eqref{eqa:26}, we obtain the optimal scheduled UE indicator $a_{m,n}^*\sq{k}$ as
\vspace{-0.3 cm}
\begin{equation}\label{eqa:28}
\begin{split}
 & a_{m,n}^*\sq{k}\\
=& \left\{ \begin{array}{l}
0, q_{m,n}^{\U}\sq{k} - q_{m,n}^{\A}\sq{k} \ge 0 \mbox{ or } q_{m,n}^{\A}\sq{k} = 0 \\
1, \mbox{otherwise}.
\end{array} \right.
\end{split}
\end{equation}

The motivation of \eqref{eqa:28} can be justified as follows.
The case $q_{m,n}^{\A}\sq{k} =0$ indicates that the backlog of the $\br{m,n}$-th access queue is zero.
The data rate $\sum\nolimits_{t_k \in \T_k} r_{m,n}\br{t_k}$ of $\br{m,n}$-th UE at the $k$-th frame is set to zero, and the $\br{m,n}$-th UE is not scheduled at the $k$-th frame.
When the case $q_{m,n}^{\U}\sq{k} - q_{m,n}^{\A}\sq{k} \ge 0$ happens, setting the value of $\sum\nolimits_{t_k \in \T_k} r_{m,n}\br{t_k}$ as zero can minimize the term \eqref{eqa:26}.
Therefore, the $\br{m,n}$-th UE is not scheduled at the $k$-th frame.

\begin{remark}
Note that the $m$-th ScBS is closed when no UE associated with the $m$-th ScBS is scheduled, i.e.,
\begin{equation}\label{eqa:28b}
\sum\limits_{n = 1}^N a_{m,n}\sq{k} = 0.
\end{equation}
\end{remark}

\subsection{Two-way Energy Trading and Beamforming Analysis}
Denote the terms related to the beamforming vectors and amount of trading energy in the RHS of \eqref{eqa:21} as
\begin{align}
{\cal OBJ}\br{t_k} =& V\br{\alpha_b - \alpha_s}\br{  P^{\SG}\br{t_k} }^+ + V\alpha_s P^{\SG}\br{t_k} \label{eqa:29}\\
&+ \summ\sum\limits_{n \in {\cal N}_m^{\act}\sq{k}} \br{q_{m,n}^{\U}\sq{k} - q_{m,n}^{\A}\sq{k}} r_{m,n}\br{t_k}.  \nonumber
\end{align}

Based on \eqref{eqa:29}, we minimize the RHS of \eqref{eqa:21} at the $k$-th frame as
\begin{align}
{\cal OPT}\sq{k} =
\min\limits_{\bar{\cal Y}\sq{k}} \;& \mathds{E}_{\cal X}\cb{ \sumtk {\cal OBJ}\br{t_k} } \label{eqa:30}\\
\st\; & \eqref{eqa:12}-\eqref{eqa:16} \nonumber
\end{align}
where $\bar{\cal Y}\sq{k} \triangleq \cb{\bm w_{m,n}\br{t_k}, P^{\SG}\br{t_k}}_{\forall m, n, t_k}$.

We observe that the challenges in solving the optimization problem \eqref{eqa:30} are three-folds:
1) the rate-limit constraints in \eqref{eqa:12} are non-convex;
2) the proportional-rate constraints in \eqref{eqa:13} are non-convex;
and 3) the optimization problem \eqref{eqa:30} contains expectation over two-scale random sources: frame-level source $\cb{E_m^{\HAV}\sq{k}}_{\forall m, k}$ and slot-level source $\cb{\bm h_{m,n}\br{t_k}}_{\forall m, n, k, t_k}$.

In order to handle the non-convex proportional-rate constraints in \eqref{eqa:13}, we introduce auxiliary variables $\cb{\phi\br{t_k}}_{\forall t_k}$ and relax the proportional-rate constraints in \eqref{eqa:13} as
\begin{align}
& \frac{\bm h^{\H}_{m,n}\br{t_k} \bm w_{m,n}\br{t_k}}{f_{m,n}\br{ \phi\br{t_k}} }    \nonumber\\
&\ge  \sqrt{  I_{m,n}^{\intra}\br{t_k} +  I_{m,n}^{\inter}\br{t_k} + \sigma_{m,n}^2 }, n \in {\cal N}_{m}^{\act}\sq{k}, \forall m \label{eqa:31}\\
& \Im\br{\bm h^{\H}_{m,n}\br{t_k} \bm w_{m,n}\br{t_k}} = 0, n \in {\cal N}_{m}^{\act}\sq{k}, \forall m  \label{eqa:32}
\end{align}
where
\begin{equation}\label{eqa:33}
f_{m,n}\br{\phi\br{t_k}}  \triangleq \sqrt{ \exp\br{ \psi_{m,n}\phi\br{t_k} } - 1 }
\end{equation}
and $\Im\br{\cdot}$ denotes the imaginary part of a complex value.
Setting the upper bound of $\phi\br{t_k}$ as $\min_{m,n}\br{{q_{m,n}^{\A}\br{t_k}}/{\psi_{m,n}}}$, the constraints in \eqref{eqa:12} are satisfied.

For fixed values of $\cb{\phi\br{t_k}}_{\forall t_k}$, the set of constraints in \eqref{eqa:31} and \eqref{eqa:32} are convex.
Substituting $r_{m,n}\br{t_k} =\psi_{m,n}\phi\br{t_k}$ and \eqref{eqa:15} into the objective function \eqref{eqa:30}, we have
\begin{equation}\label{eqa:34}
\begin{split}
& \overline{\cal OBJ}\br{t_k} = \\
& V   \br{\alpha_b - \alpha_s}\br{ \summ \br{ P_m^{\SC}\br{t_k} - \frac{1}{T} E_m^{\HAV}\sq{k}} }^+   \\
& + V\alpha_s \summ  \br{  P_m^{\SC}\br{t_k} - \frac{1}{T} E_m^{\HAV}\sq{k}}  \\
& + \summ\sum\limits_{n \in {\cal N}_m^{\act}\sq{k}}\br{ q_{m,n}^{\U}\sq{k} - q_{m,n}^{\A}\sq{k}}\psi_{m,n} \phi\br{t_k}.
\end{split}
\end{equation}

Replacing constraints in \eqref{eqa:12} and \eqref{eqa:13} with \eqref{eqa:31} and \eqref{eqa:32}, we obtain a relaxed version of optimization problem \eqref{eqa:30} as
\begin{align}
 \overline{\cal OPT}\sq{k} =
\min\limits_{\widetilde{\cal Y}} \;& \mathds{E}_{\cal X}\cb{  \sumtk \overline{\cal OBJ}\br{t_k} } \label{eqa:35}\\
\st\; & \eqref{eqa:14}, \eqref{eqa:31}, \eqref{eqa:32} \nonumber
\end{align}
where $\widetilde{\cal Y} \triangleq \cb{ \bm w_{m,n}\br{t_k} }_{\forall m, n, k, t_k}$.

Given the optimal $\phi^*\br{t_k}$, the constraints in \eqref{eqa:31} and \eqref{eqa:32} constitute a convex hull of the constraints in \eqref{eqa:13}.
Hence, we conclude that $\overline{\cal OPT}\sq{k} \le {\cal OPT}\sq{k}$.
With slight modification of the arguments in \cite{DongJan.2019}, we demonstrate that the optimal beamforming vectors $\cb{\bm w_{m,n}^*\br{t_k}}_{\forall m, n, k, t_k}$ make the constraints in \eqref{eqa:31} active.
In other words, we have $\overline{\cal OPT}\sq{k} = {\cal OPT}\sq{k}$.
The detailed proof of the activeness of \eqref{eqa:31} will be provided in the extended version of the conference article.
Motivated by \cite{DongJan.2019}, the optimal $\phi^*\br{t_k}$ can be obtained via a one-dimensional search method.

Now the optimization problem \eqref{eqa:35} is convex with respect to $\widetilde{\cal Y}$.
Let $E_m^{\HAV}\sq{k}$ and $\cb{\bm h_{m,n}\br{t_k}}_{\forall m,n}$ denote the amount of harvested NRE at the $m$-th ScBS at the \mbox{$k$-th} frame  and the set of random sources at the $t_k$-th slot.
Since the optimization problem \eqref{eqa:35} contains two-scale random sources, we are motivated to use the principle of opportunistically minimizing an expectation \cite{Neelybook} with the assumption that the channel coefficient vectors $\cb{\bm h_{m,n}\br{t_k}}_{\forall m, n}$ are independent and identically distributed (i.i.d.) over different slots.

Based on the aforementioned discussions, we summarize the procedures of our proposed joint beamforming, UE scheduling and ScBS sleeping algorithm as follows.

\vspace{-0.3 cm}
\begin{algorithm}[ht]\small
  \centering
  \caption{Joint Beamforming, UE Scheduling and ScBS Sleeping Algorithm}\label{alg:01}
  \begin{algorithmic}[1]
  \State At the start of the $k$-th frame, the CN estimates the harvested NRE as $\cb{E_m^{\HAV}\sq{k}}_{\forall m}$
  \State At the start of the $k$-th frame, the CN updates the set of scheduled UEs and active ScBSs via \eqref{eqa:28} and \eqref{eqa:28b}
  \Repeat
  \State At the start of the $t_k$-th slot, the CN estimates the channel coefficient vector $\cb{\bm h_{m,n}\br{t_k}}_{\forall m, n}$
  \State With  $\cb{\bm h_{m,n}\br{t_k}}_{\forall m, n}$, the CN solves the following optimization problem via CVX \cite{Grant2014}
  \begin{equation}
  \begin{split}
    \min\limits_{\widetilde{\cal Y}} \;&  \overline{\cal OBJ}\br{t_k} \\
    \st\; & \eqref{eqa:14}, \eqref{eqa:31}, \eqref{eqa:32}
  \end{split}
  \end{equation}
  \State At the start of the $t_k$-th slot, the CN performs one dimensional search for the optimal $\phi^*\br{t_k}$
  \Until{Convergence}
  \State At the end of the $t_k$-th slot, the CN updates the access queues and processing queues according to \eqref{eqa:02} and \eqref{eqa:04}
  \end{algorithmic}
\end{algorithm}

\begin{proposition}\label{pr:02}
Suppose the arrival rate and service rate of the $\br{m,n}$-th access queue and the $\br{m,n}$-th processing queue satisfy the condition $\frac{1}{T}\lambda_{m,n} <  \mathds{E}_{\cal X}\cb{r_{m,n}\br{t_k}} <  \bar s_{m,n}$.
When the proposed joint beamforming, UE scheduling and ScBS sleeping algorithm is used, we conclude that
\begin{itemize}
  \item The optimal grid-energy expenditure is asymptotically obtained as
    \begin{equation}
    G^* \le \frac{1}{K}\sumk \mathds{E}_{\cal X}\cb{ G\sq{k} } \le \frac{\Psi}{V} + G^*
    \end{equation}
    where $G^*$ is the optimal grid-energy expenditure.
  \item The constraints in \eqref{eqa:16} are satisfied.
\end{itemize}
\end{proposition}

\begin{IEEEproof}
Due to the space limitation, the detailed proof is omitted and it will be provided in the extended version of this conference paper.
\end{IEEEproof}

\begin{figure*}[bt]
\vspace{-0.3 cm}
\setcounter{equation}{36}
\begin{equation}\label{apdx:02:08}
\begin{split}
& \Delta_{\X} + \mathds{E}_{\cal X}\cb{ G\sq{k} }  \le \Psi + \mathds{E}_{\cal X}\cb{ G\sq{k} }\\
& + \summ\sumn q_{m,n}^{\A}\sq{k} \mathds{E}_{\cal X}\cb{ \lambda_{m,n} - \sumtk r_{m,n}\br{t_k}  } + \sumn\sumn q_{m,n}^{\U}\sq{k} \mathds{E}_{\cal X}\cb{ \sumtk\br{r_{m,n}\br{t_k} - s_{m,n}\br{t_k}} }.
\end{split}
\end{equation}
\vspace{-0.5 cm}
\hrulefill
\end{figure*}

\vspace{-0.2 cm}
\section{Numerical Results}
In the section, we use numerical results to verify our proposed algorithm.
We perform the simulations based on practical data from NASA Remote Sensing Validation Data in Saudi Arabia: Solar Valley at 10:00 am to 10:06 am, on December 2000\footnote{$\mbox{https://www.nrel.gov/grid/solar-resource/saudi-arabia.html}$}.
Each ScBS is equipped with an solar energy harvester with size $5 \mbox{ cm}^2$ and   harvesting efficiency 30\%.
Since NRE arrival rate is updated every 5 minutes, we use the interpolation method to generate NRE arrival rate at 0.5 second level.
Since the electricity prices vary every hour, the purchasing price and selling price are, respectively, set as $1.2 \times 10^{-9}$ cents/slot/mW and $1 \times 10^{-9}$ cents/slot/mW based on the practical data of Pennsylvania-New~Jersey-Maryland market\footnote{$\mbox{https://www.pjm.com/markets-and-operations.aspx}$}.
We consider wireless UE has a speed at $v_{\mbox{\tiny UE}} = 1.5$ km/h \cite{3gpp_ue_speed}.
The Doppler frequency shift $f_{\mbox{\tiny D}} = \frac{v_{\mbox{\tiny UE}}}{C} f_{\mbox{\tiny C}} = 1.25$ Hz, where the light speed $C$ is $3\times 10^8$ m/sec.
In order to guarantee a slow-fading scenario, we choose the slot duration as  $0.1$ sec \cite{3gpp_ue_speed}.
The processing rate at the UEs is $s_{m,n} = 3.5$ nats/slot/Hz.
The power of AWGN is set as $-90$ dBm.
The maximum transmit power of ScBS is set as $P_m^{\max} = 26$ dBm.
The consumed power on baseband processing is set as $P_m^{\SP} = 23$ dBm.

\vspace{-0.3 cm}
\begin{figure}[ht]
\centering
\includegraphics[width=2.6 in]{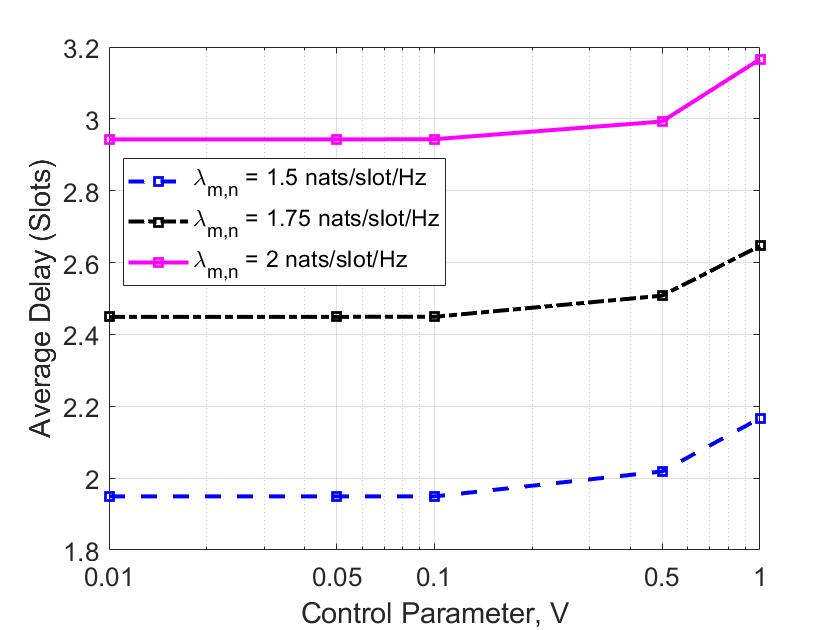}
\caption{Variation of the average delay with the control parameter.}\label{fg:01}
\includegraphics[width=2.6 in]{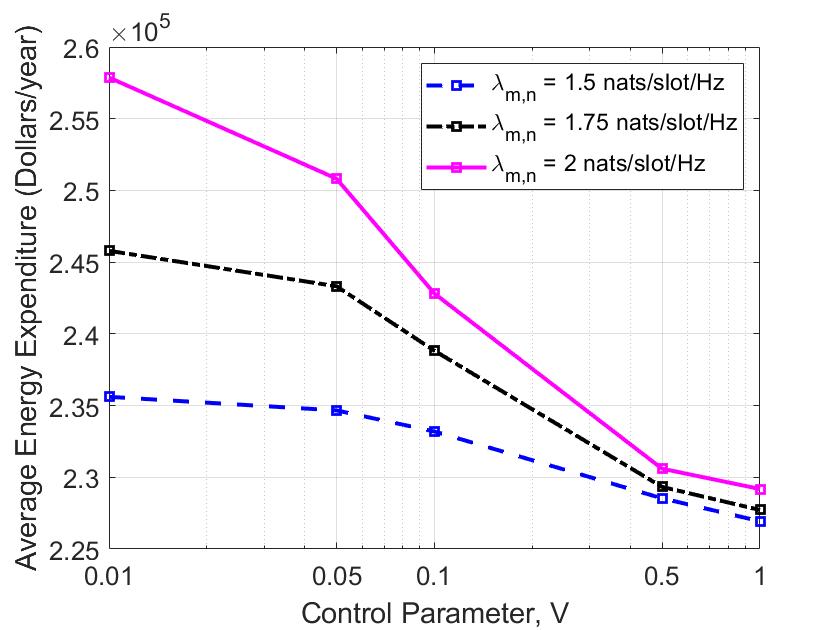}
\caption{Variation of the annualized electricity expenditure with the control parameter.}\label{fg:02}
\end{figure}
\vspace{-0.3 cm}

Figures \ref{fg:01} and \ref{fg:02} reveal the tradeoff between the average end-to-end delay and grid-energy expenditure.
Here, we annualize the grid-energy expenditure by considering the medium city with $1\times 10^5$ ScBSs.
We observe that increasing the value of control parameter induces an increasing end-to-end delay of UEs (as shown in Fig. \ref{fg:01}) and a decreasing grid-energy expenditure (as shown in Fig. \ref{fg:02}).
Therefore, the proposed algorithm provides the wireless operator with flexibility in controlling the grid-energy expenditure while maintaining a satisfactory level of communication QoS.
Moreover, increasing the information arrival rate of UEs induces a more efficient tradeoff between the end-to-end delay and grid-energy expenditure.
For example, when the information arrival rate $\lambda_{m,n}$ is $1.5$ nats/slot/Hz, the wireless operator can trade the $0.21$ slot of end-to-end delay for $3.68$\% grid-energy expenditure by increasing the value of control parameter from $0.01$ to $1$.
When the information arrival rate $\lambda_{m,n}$ is $2$ nats/slot/Hz, the wireless operator can trade the $0.22$ slot of end-to-end delay for $11.13$\% grid-energy expenditure with the same range of control parameter.
This is due to the three facts: 1) the increasing control parameter induces that the grid-energy expenditure approaches the same optimal value with different information arrival rates; 2) a smaller control parameter indicates a more stringent delay requirement; and 3) a larger stringent delay requirement exponentially increases the grid-energy expenditure due to the log-concave data rate of UEs.

\vspace{-0.3 cm}
\section{Conclusion}
We developed a joint beamforming, UE scheduling and ScBS sleeping algorithm.
Within the proposed algorithm, scheduling UEs and switching ScBSs are performed at the coarse-grained granularity (frame) while calculating beamforming vectors are respectively performed at the fine-grained granularity (slot).
The benefit of the two scale algorithm is to avoid frequently changing scheduled UEs and switching on/off the ScBSs.
Numerical results demonstrate that the wireless operator can trade end-to-end delay for grid-energy expenditure by tuning a control parameter.
Moreover, increasing the maximum transmit power of ScBSs can improve the tradeoff efficiency between end-to-end delay and grid-energy expenditure.

\vspace{-0.3 cm}
\appendix
\setcounter{equation}{33}
\section{Proof of Proposition \ref{pr:01}}\label{apdx:01}
Taking the telescoping summation over $kT \le t_k < \br{k+1}T - 1$ for $q_{m,n}^{\A}\br{t_k}$ in \eqref{eqa:02}, we obtain the frame-by-frame dynamic equation of the $\br{m,n}$-th access queue as
\begin{equation}\label{apdx:02:01}
q_{m,n}^{\A}\sq{k+1} = q_{m,n}^{\A}\sq{k}  + \lambda_{m,n} - \sumtk r_{m,n}\br{t_k}.
\end{equation}
Based on \eqref{apdx:02:01}, the one-frame drift of the $\br{m,n}$-th access queue is denoted as
\begin{equation}\label{apdx:02:02}
\begin{split}
 & \frac{1}{2}\br{ q_{m,n}^{\A}\sq{k+1} }^2 - \frac{1}{2}\br{ q_{m,n}^{\A}\sq{k} }^2 \\
\le &  \C^{\A}_{m, n} + q_{m,n}^{\A}\sq{k} \br{ \lambda_{m,n} - \sumtk r_{m,n}\br{t_k}  }
\end{split}
\end{equation}
where $\C^{\A}_{m, n} \triangleq \frac{\lambda_{m,n}^2 + T^2 \br{r_{m,n}^{\max}}^2 }{2}$ due to the fact that $r_{m,n}\br{t_k} \in \sq{0, r_{m,n}^{\max}}$.

Following a similar argument, we obtain the one-frame drift of the $\br{m,n}$-th processing queue as
\begin{equation}\label{apdx:02:04}
\begin{split}
& \frac{1}{2}\br{ q_{m,n}^{\U}\sq{k+1} }^2 - \frac{1}{2}\br{ q_{m,n}^{\U}\sq{k} }^2 \\
\le & \C^{\U}_{m,n} + q_{m,n}^{\U}\sq{k}\sumtk\br{  r_{m,n}\br{t_k} - s_{m,n}\br{t_k} }
\end{split}
\end{equation}
where $\C^{\U}_{m,n} \triangleq T^2\frac{\bar s_{m,n}^2 + \br{r_{m,n}^{\max}}^2}{2}$.

Let $\Psi \triangleq \sum\nolimits_{m=1}^M\sum\nolimits_{n =1}^{N_m}\br{ \C_{m,n}^{\A} + \C_{m,n}^{\U} }$.
Based on \eqref{apdx:02:02} and \eqref{apdx:02:04}, the one-frame Lyapunov drift-plus-penalty function in \eqref{eqa:20} is derived as in \eqref{apdx:02:08}.

\vspace{-0.3 cm}
\bibliographystyle{IEEEtran}
\bibliography{dyj_bib}
\end{document}